\documentclass[
   final            
  ]
  {aipproc}

\layoutstyle{6x9}

\begin{document}

\title{Multi Jet Production at High $Q^2$}


\author{Thomas Kluge \footnote{on behalf of the H1 Collaboration}}{
  address={DESY, Notkestr. 85, 22607 Hamburg, Germany\\E-mail: thomas.kluge@desy.de}
}



\begin{abstract}
Deep-inelastic $e^+p$ scattering data, taken with the H1 detector at HERA, are used to investigate jet production over a range of four-momentum transfers $150 < Q^2 < 15000 \,\mathrm{GeV}^2$ and transverse jet energies $5 < E_T < 50 \,\mathrm{GeV}$.
The analysis is based on data corresponding to an integrated luminosity of $\mathcal{L}_\mathrm{int} = 65.4\,\mathrm{pb}^{-1}$ taken in the years 1999-2000 at a centre-of-mass energy $\sqrt{s} \approx 319\,\mathrm{GeV}$.
Jets are defined by the inclusive $k_t$ algorithm in the Breit frame of reference.
Dijet and trijet jet cross sections are measured with respect to the exchanged boson virtuality and in addition the ratio of the trijet to the dijet cross section $R_{3/2}$ is investigated.
The results are compared to the predictions of perturbative QCD calculations in next-to-leading order in the strong coupling constant $\alpha_s$.
The value of $\alpha_s(m_Z)$ determined from the study of $R_{3/2}$ is $\alpha_s(m_Z) = 0.1175 \pm 0.0017 (\mathrm{stat.}) \pm 0.0050 (\mathrm{syst.}) ^{+0.0054}_{-0.0068} (\mathrm{theo.})$.
\end{abstract}

\maketitle


\section{Introduction and Observables}
Deep-inelastic scattering (DIS) at HERA is a precision tool for studies of Quantum Chromo Dynamics (QCD).
Jet cross sections at high transverse momenta are in particular attractive, because perturbative calculations (pQCD) are precise and non-perturbative hadronisation effects are weak.
In the past inclusive jet cross sections \cite{Adloff:2000tq,Chekanov:2002be}, as well as multijet cross sections \cite{Adloff:2001kg,Chekanov:2005ve} have been studied at HERA.

The aim of the present analysis is to check quantitatively pQCD predictions for di- and trijet cross sections and to determine the value of the strong coupling at different values of the relevant hard scale.

In the following, jets are defined by the inclusive $k_t$ cluster algorithm \cite{Ellis:1993tq,Catani:1993hr} in the Breit frame of reference.
Accepted jets are required to have a transverse energy of more than  $5\,\mathrm{GeV}$. 
These jets, boosted back to the laboratory frame, have to fulfill the pseudorapidity requirement $-1<\eta_\mathrm{lab}<2.5$ to ensure good detector acceptance.
Events with at least two (three) accepted jets are assigned to the dijet (trijet) event sample.
In order to ensure stability of the perturbative calculations, cuts on the invariant jet masses are applied:  $M_\mathrm{dijet}>25\,\mathrm{GeV}$ and  $M_\mathrm{trijet}>25\,\mathrm{GeV}$ for the di- and trijet sample, respectively.

\section{Data Sample and Analysis Methods}
The data this analysis is based on were taken with the H1 detector in the years 1999-2000 and correspond to an integrated luminosity of $\mathcal{L}_\mathrm{int} = 65.4\,\mathrm{pb}^{-1}$.
A DIS selection in the phase space of four-momentum transfers $150 < Q^2 < 15000 \,\mathrm{GeV}^2$ and inelasticity $0.2<y<0.6$ leaves
5460 dijet- and 1757 trijet events.

To account for limited detector acceptance and resolution, correction factors were applied, which were determined with the Monte Carlo event generators DJANGOH \cite{Charchula:1994kf} (including ARIADNE \cite{Lonnblad:1992tz}) and RAPGAP \cite{Jung:1995gf}.
The cross sections were corrected for QED radiative effects with HERACLES \cite{Kwiatkowski:1992es}.

Relevant experimental uncertainties include: the electromagnetic energy scale, the hadronic energy scale, the scattering angle of the positron, a model uncertainty in the acceptance correction and the luminosity measurement, with the hadronic energy scale uncertainty being the dominant contribution.

\section{Results}
\begin{figure}
  \includegraphics[width=.8\textwidth]{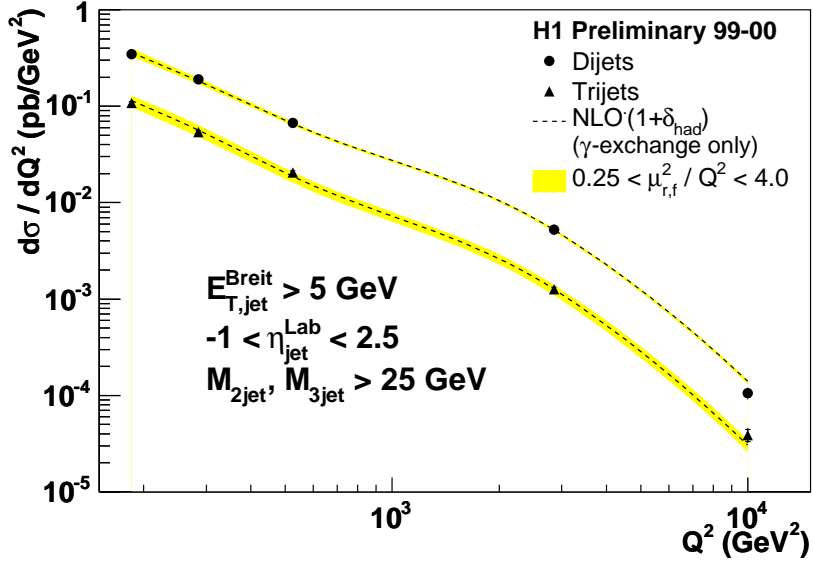}
  \label{fig1}
  \caption{NC dijet and trijet differential cross-sections, with respect to $Q^2$, shown with NLO pQCD predictions including hadronisation corrections. The shaded bands show the effect of varying the renormalisation/factorisation scale by a factor of two.}
\end{figure}

Fig.~\ref{fig1} shows the differential di- and trijet cross sections as a function of $Q^2$.
The data are compared to a calculation carried out with NLOJET++ \cite{Nagy:2001xb}, using parton density functions of the proton from the CTEQ5M1 \cite{Lai:1999wy} set  and assuming a value of the strong coupling $\alpha_s(m_Z) = 0.118$.
The calculation includes matrix elements at next-to-leading order (NLO) for two and three partons in the final state, corresponding
to orders of the strong coupling up to $\mathcal{O}(\alpha_s^3)$. 
Hadronisation corrections are applied to the parton level prediction, determined with the help of DJANGOH and RAPGAP event samples.
An uncertainty estimate for the prediction is provided by variations of the renormalisation and factorisation scale by a factor of two.
The cross sections span several orders of magnitude, where a  good description of the data by the prediction is observed.
At the highest $Q^2$ bin electroweak effects, which are not present in the calculation, cannot be neglected.
Hence this point will not be used in the following $\alpha_s$ fit.

\begin{figure}
  \includegraphics[width=.8\textwidth]{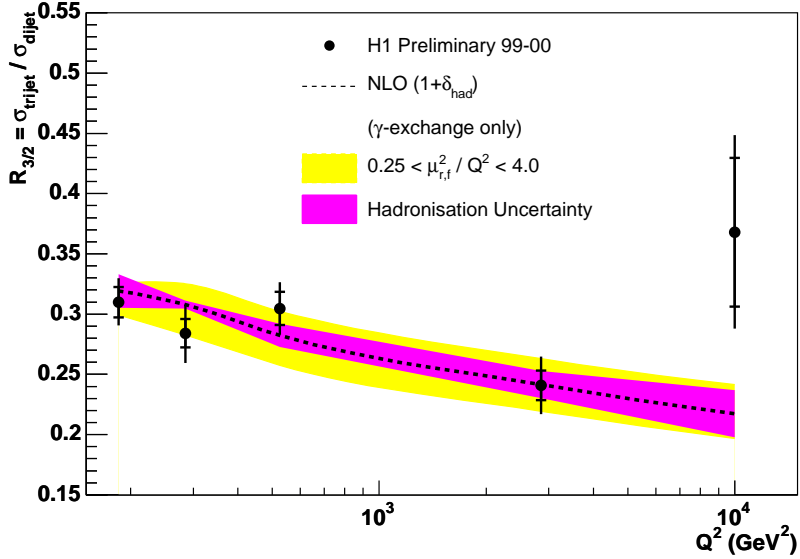}
  \label{fig2}
  \caption{Measured values of $R_{3/2}$ against $Q^2$ compared with a NLO pQCD prediction with hadronisation corrections. The dark shaded band shows the uncertainty associated with the hadronisation corrections, while the light shaded band shows the effect of varying the renormalisation/factorisation scale by a factor of two. }
\end{figure}

The ratio of the tri- and dijet cross section, $R_{3/2}=\sigma_\mathrm{trijet}/\sigma_\mathrm{dijet}$ is shown in Fig.~\ref{fig2}, where again a good description by the perturbative calculation is observed.
Based on this measurement, a fit of the strong coupling is performed: 
for each $Q^2$ value at which $R_{3/2}$ is measured (but the highest one) the pQCD calculation is repeated with the five variations of the proton p.d.f.s, available within the CTEQ4A \cite{Lai:1996mg} set.
This yields five predictions of $R_{3/2}$ as a function of $\alpha_s(m_Z)$,
to which a function of the form $C_1\alpha_s(m_Z)+C_2\alpha_s^2(m_Z)$  is fitted.
Consequently, the function is used to obtain from the measured value of $R_{3/2}$ and its uncertainty the corresponding value and uncertainty of $\alpha_s(m_Z)$.

\begin{figure}
  \includegraphics[width=.8\textwidth]{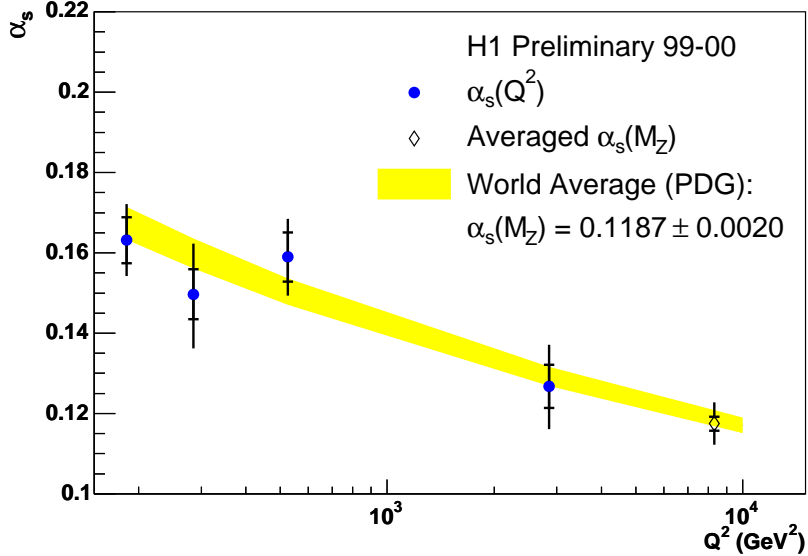}
  \label{fig3}
  \caption{$\alpha_s(m_Z)$ values from each $Q^2$ bin evolved to values at their respective values of $Q^2$ (points) using the two-loop solution of the renormalisation group equation. The averaged value of $\alpha_s(m_Z)$, found using a $\chi^2$ minimisation fit, is shown at the far right (empty diamond).  The inner error bars show the statistical errors alone and the outer error bars denote the quadratic sum of statistical and systematical errors. The evolution of the world average value of $\alpha_s(m_Z)$ is shown as a shaded band. }
\end{figure}

The fit results are shown as points in Fig.~\ref{fig3}.
From this diagram the running of the renormalised strong coupling is clearly evident.
In addition an average value is obtained from this values by $\chi^2$ minimisation, shown as a diamond at the reference of $Q^2=m_Z^2$.
The presented result is well compatible with the world average from the PDG \cite{Eidelman:2004wy}, which  is shown  as a band for comparison.  

\section{Conclusion}
Differential di- and trijet cross sections measured at high $Q^2$ have been shown.
The distributions are well described by NLO pQCD with hadronisation corrections, except for the highest $Q^2$ data point, where electroweak corrections cannot be neglected.   
A fit of the strong coupling yields \begin{center}$\alpha_s(m_Z) = 0.1175 \pm 0.0017 (\mathrm{stat.}) \pm 0.0050 (\mathrm{syst.}) ^{+0.0054}_{-0.0068} (\mathrm{theo.}),$\end{center} well in agreement with the world average.
Future improvements in the understanding of the hadronic energy scale will significantly reduce the experimental uncertainty.



\bibliographystyle{aipproc}   

\bibliography{tkbib}

\IfFileExists{\jobname.bbl}{}
 {\typeout{}
  \typeout{******************************************}
  \typeout{** Please run "bibtex \jobname" to optain}
  \typeout{** the bibliography and then re-run LaTeX}
  \typeout{** twice to fix the references!}
  \typeout{******************************************}
  \typeout{}
 }

\end{document}